\documentclass{article}
\usepackage[a4paper, left=1.5cm, right=1.5cm, top=2.5cm, bottom=2.5cm]{geometry}
\usepackage{graphicx} % Required for inserting images
\usepackage{amsmath}
\usepackage{amssymb}
\usepackage{xcolor}
\usepackage{authblk} % For author and affiliation formatting
\usepackage{cite}

\title{Quantitative Mapping of the Loeb Scale}

\author[1]{Oem Trivedi\thanks{oem.trivedi@vanderbilt.edu}}
\author[2]{Abraham Loeb\thanks{aloeb@cfa.harvard.edu}}

\affil[1]{Department of Physics and Astronomy, Vanderbilt University, Nashville, TN 37235, USA}
\affil[2]{Astronomy Department, Harvard University, 60 Garden St., Cambridge, MA 02138, USA}

\date{\today}

\begin{document}

\maketitle

\begin{abstract}
    The recent discovery of a third interstellar object (ISO) 3I/ATLAS, following 1I/`Oumuamua and 2I/Borisov, has raised questions about the nature and origin of these enigmatic objects. With the Vera C. Rubin Observatory expected to discover dozens of new ISOs over the next decade, it is timely to use a classification scheme for their nature in the context of the recently proposed Loeb scale. Here, we provide a formalism for ranking ISOs quantitatively on the Loeb Scale in analogy to Sagan's formalism for mapping the Kardashev scale based on the energy output of technological civilizations.   
\end{abstract}

\section{Introduction}
Over the past decade, three interstellar objects (ISOs) have been identified by astronomical observations: 1I/`Oumuamua in 2017 \cite{oumumeech2017brief}, 2I/Borisov \cite{borguzik2020initial} in 2019, and 3I/ATLAS in 2025 \cite{seligman2025discovery}. These discoveries have opened up a new observational window into the analysis of large physical bodies from outside the solar system. While 2I/Borisov appeared as a conventional comet, both 1I/`Oumuamua and 3I/ATLAS exhibited anomalous properties that deviate from familiar solar system objects\cite{oum1loeb2022possibility,oum2forbes2019turning,oum3siraj2022mass,oum4siraj20192019,oum5bialy2018could,3i1loeb20253i,3i2a2hibberd2025interstellar,3i3tde2025assessing,3i4a3loeb2025intercepting,3i5hopkins2025different}. With the advent of the Vera C. Rubin Observatory \cite{vc1thomas2020vera,vc2blum2022snowmass2021,vc3sebag2020vera}, it is anticipated that the detection rate of interstellar objects will increase by up to two orders of magnitude, raising the need for a systematic framework to classify and interpret their nature and distinguish between natural icy rocks and possible technological artifacts. 
\\
\\
The Loeb Scale was recently proposed as such a framework \cite{eldadi2025loeb} as it provides a structured, ten-level classification scheme to evaluate ISOs based on the level of anomalies that they exhibit relative to icy rocks, raising the possibility that they might be artificial in origin. Much like the Kardashev scale offers a way to classify civilizations according to their energy usage \cite{kar1kardashev1964transmission,kar2kardashev1980strategies,kar3cirkovic2016kardashev,kar4gray2020extended,kar5namboodiripad2021predicting,kar6sonia2022civilizations}, the Loeb Scale serves as a tool for ranking interstellar artifacts along a continuum ranging from ordinary natural bodies (Level 0) to confirmed technological artifacts that potentially pose a risk to humanity (Levels 8-10). Whereas the Loeb Scale offers a systematic classification scheme, it does not apply to the civilizations themselves but rather to their products based on observational surveys near Earth. 
\\
\\
Carl Sagan refined the Kardashev framework by introducing a logarithmic interpolation scheme that relates a civilization type to its total energy consumption \cite{sagan1973detectivity}, thereby enabling a continuous quantitative mapping as a substitute to a coarse categorical distinction. Here, we apply a similar approach to the Loeb Scale for ISOs by mapping their measured anomalies and physical characteristics to a quantitative rank. The goal of this paper is to refine the Loeb Scale by translating observable features of ISOs into a continuous score. In section II we provide a brief overview of the Loeb scale, and in section III we explictly define this quantitative mapping. In section IV we apply this mapping to derive the Loeb scale rankings for the three known ISOs. Finally, we summarize oiur main conclusions in section V.
\\
\\
\section{The Loeb Scale}
The Loeb scale is a classification system, designed to quantify the technological significance of anomalous properties of ISOs. Motivated by the anomalies of 1I/‘Oumuamua and 3I/ATLAS, the scale provides a structured way to place new discoveries within an organized hierarchy that ranges from purely natural origins to confirmed manifestations of extraterrestrial technology. Its purpose is to offer the astronomical community a framework for discussing technological markers on ISOs \cite{Daven25}, in the same way that the Kardashev scale provides a framework for categorizing technological civilizations \cite{kar1kardashev1964transmission,kar2kardashev1980strategies,kar3cirkovic2016kardashev,kar4gray2020extended,kar5namboodiripad2021predicting,kar6sonia2022civilizations}. 
\\
\\
The Loeb scale begins at level 0, corresponding to natural astrophysical bodies whose properties are well understood and require no appeal to exotic explanations, and rises to level 10, reserved for catastrophic scenarios involving confirmed technological objects that pose an existential threat to humanity on Earth. At the lower levels, the scale is meant to separate trivial or explainable anomalies from those requiring closer scrutiny, with Level 0 referring to ISOs that exhibit properties consistent with known classes of comets or asteroids, such as the interstellar comet 2I/Borisov. Level 1 captures objects with minor deviations or inconclusive data that do not strongly challenge a natural interpretation, whereas level 2 describes an ISO that exhibits some unusual properties but still falls within the range of natural phenomena given current understanding. Level 3 marks the point where persistent anomalies appear, with features such as repeated but unexplained variations in trajectory or unusual albedo or morphology, without yet triggering serious consideration of artificial origin.
\\
\\
The middle of the scale is where the possibility of technosignatures is formally acknowledged, where Level 4 is described as the “critical threshold,” where the combination of anomalies is strong enough that a technological explanation must be considered alongside natural hypotheses. This was the classification assigned to 1I/‘Oumuamua, whose non-gravitational acceleration without a coma of gas or dust, spectral properties, and extreme shape remain unexplained by conventional models. Level 5 corresponds to a suspected passive technology, such as derelict probes, light sails, or debris, where evidence suggests artificial origin but no active behavior is detected. Levels 6 and 7 escalate to suspected active technologies, including evidence of propulsion, maneuvering, or directed electromagnetic emissions. At these levels, the possibility of interaction or intent must be contemplated but the distinction between minor and major threats to humanity can be ignored.
\\
\\
The uppermost part of the scale deals with confirmation and risk, with Level 8 being assigned when multiple lines of evidence establish beyond a reasonable doubt that the object is artificial but it poses no immediate danger, for example a flyby of a confirmed probe near Earth. Level 9 is reached if the confirmed artificial object is on a trajectory with the potential to cause regional harm, such as an impact equivalent to a major nuclear strike. Finally, Level 10 is the highest designation and is reserved for a confirmed artificial body on a collision course with Earth that would carry global consequences, akin to existential threats in planetary defense assessments. By laying out this graded sequence from the mundane to the catastrophic, the Loeb scale provides both a scientific framework for classifying ISOs and a practical guide for allocating observational and policy attention in proportion to the level of the risk.
\\
\\
\section{Mapping the Scale}
Motivated by Sagan's \cite{sagan1973detectivity} mapping of civilization power to the Kardashev type \cite{kar1kardashev1964transmission}, which maps a continuous measure of the total energy use to a discrete scale of capability, we seek a quantitative mapping from measured ISO properties to a single continuous rank on the Loeb Scale. The design goals are reproducibility, interpretability (so the community can see which observables drive the classification), tunability (constants and weights can be adjusted as the sample of ISOs grows), and hard trigger override rules that reflect  the prescription in \cite{eldadi2025loeb} in which certain observations should immediately escalate the classification.
\\
\\
We first define a set of normalized observable metrics, each constrained to the interval $[0,1]$ where $0$ denotes behavior fully consistent with natural expectations and $1$ denotes behavior that is maximally anomalous or maximally suggestive of technology. The metrics are:
\begin{align*}
A &: \text{non-gravitational acceleration anomaly score},\\
B &: \text{spectral / composition anomaly score},\\
C &: \text{shape / lightcurve anomaly score},\\
D &: \text{albedo / surface-weathering anomaly score},\\
E &: \text{trajectory / targeting improbability score},\\
F &: \text{electromagnetic (EM) signal score},\\
G &: \text{operational / behaviour score (maneuvering, sub-objects)},\\
H &: \text{impact-risk factor (optional, can be used for Levels 8-10)}.
\end{align*}
Each metric $M\in{A,B,\dots,G}$ is constructed from measured quantities and then normalized into $[0,1]$ by application of monotonic transforms and clamping based on the functional forms given below. We note that the constants appearing in these transforms are tunable calibration parameters.
\\
\\
The non-gravitational acceleration anomaly raw value is defined by the ratio of observed non-gravitational acceleration $a_{\rm obs}$ to a reference typical cometary acceleration $a_{\rm ref}$
\begin{equation}
    A_{\rm raw}=\log_{10}\!\left(\frac{a_{\rm obs}}{a_{\rm ref}}\right).
\end{equation}
A convenient normalization that compresses a wide dynamic range into $[0,1]$ is
\begin{equation}
    A=\operatorname{clamp}\!\left(\frac{A_{raw}+2}{4},0,1\right),
\end{equation}
where Clamp$(x,0,1)=\min(\max(x,0),1)$ and $a_{\rm ref}$ is chosen by some community calibration (a nominal choice could be $a_{\rm ref}=10^{-6}~\mathrm{m~s^{-2}}$). The constants $+2$ and denominator $4$ set the mapping so that objects with $a_{\rm obs}\approx a_{\rm ref}$ yield $A\approx 0.5$.
\\
\\
The spectral or composition anomaly $B$ is constructed by quantitative comparison to a library of natural spectra. If we let $\chi^2_{ mismatch}$ represent a goodness of fit measure between the observed spectrum and the best-fit natural taxonomy model then one can normalize this via
\begin{equation}    B=\operatorname{clamp}\!\left(\frac{\chi^2_{mismatch}}{\chi^2_{ mismatch}+K_B},0,1\right),
\end{equation}
where $K_B>0$ is a calibration constant controlling sensitivity to spectral mismatch. It is important to discuss some subtleties of this parametrization. The spectral anomaly metric $B$ encodes whether the observed spectral and compositional signatures of an ISO are consistent with those of Solar System comets or asteroids. Rather than assigning arbitrary anomaly levels, we compare the measured gas production rates, line ratios, and gas-to-dust quantities to the known population distributions compiled in cometary taxonomic studies. For a measured value $x^\star$, we define its population percentile as $F_{\rm pop}(x^\star)$ and assign a rarity score
\begin{equation}
    s_x = 1 - 2\min\!\big(F_{\rm pop}(x^\star),1-F_{\rm pop}(x^\star)\big)
\end{equation}
For censored data (upper limits, such as the non-detection of CN emission in 3I/ATLAS \cite{cnemissions}), we instead use the population survival function
\begin{equation}
    s_x = 1 - 2\min\!\left(\int_{x_{\rm UL}}^\infty p_{\rm pop}(x)dx,1-\int_{x_{\rm UL}}^\infty p_{\rm pop}(x)dx\right)
\end{equation}
so that unusually stringent upper limits contribute properly to the anomaly score. Multiple spectral indicators are then combined as a weighted sum of their rarity scores and transformed into the metric
$$
B = \operatorname{clamp}\!\left(\frac{\sum_k \alpha_k s_{x_k}}{\sum_k \alpha_k s_{x_k}+K_B},0,1\right),
$$
with equal weights $\alpha_k$ in the present examples and $K_B=1$.
Applying this procedure, 2I/Borisov shows a standard cometary spectrum with strong CN, C$_2$, and C$_3$ bands and dust consistent with Solar System comets, producing $B\approx 0.00$. For 1I/`Oumuamua, no gas or dust was detected despite a non-gravitational acceleration, an unusual combination that places it far from the Solar System comet distributions and this yields a high rarity score and $B\approx 0.80$. For 3I/ATLAS, the CN production onset at $\sim3.1$ au was unusually early, the object was among the most carbon-chain depleted known with $\log(Q(\mathrm{C}_2)/Q(\mathrm{CN}))<-1.05$, and the gas-to-dust ratio was at the low end of the Solar System distribution. These indicators jointly produce a rarity score $s_x\sim0.8$, resulting in $B\approx 0.80$.
\\
\\
The shape/lightcurve anomaly $C$ is based on an inferred aspect ratio $R$ or on specialized lightcurve-model residuals and a simple mapping is
\begin{equation}    C=\operatorname{clamp}\!\left(\frac{\log_{10}(R)}{\log_{10}(R_{\max})},0,1\right),
\end{equation}
with $R_{\max}$ a reference extreme aspect ratio (for example $R_{\max}=10$). Under this mapping $R\ge R_{\max}\Rightarrow C\approx 1$.
\\
\\
The albedo or surface-weathering anomaly $D$ is produced by comparing the measured geometric albedo, color gradients, and cosmic-ray exposure indicators to expectations given inferred travel time and cosmic-ray fluence. If $s_{albedo}$ is a normalized discrepancy measure then we have
\begin{equation}    D=\operatorname{clamp}\!\left(\frac{s_{albedo}}{s_{ albedo}+K_D},0,1\right),
\end{equation}
with calibration constant $K_D$. We note here that albedo anomaly metric is defined relative to the known distribution of asteroid and comet albedos rather than by an ad hoc discrepancy measure. To do this, we adopt the two-Rayleigh mixture distribution for the geometric albedo $p_V$ derived from NEOWISE data \cite{wright2016albedo}, which is given by
\begin{equation}
    p_{2R}(p_V) = f_D {\rm Ray}(p_V;d) + (1-f_D){\rm Ray}(p_V;b),
\end{equation}
with Rayleigh components
\begin{equation}
    {\rm Ray}(x;\sigma) = \frac{x}{\sigma^2}\exp\!\left(-\frac{x^2}{2\sigma^2}\right), \qquad x\ge0 ,
\end{equation}
and fitted parameters
\begin{equation}
    f_D=0.253,\qquad d=0.030,\qquad b=0.168.
\end{equation}
This mixture accurately represents the bimodal albedo distribution of small bodies in the Solar System. Given an observed albedo $p_V^\star$, we compute its tail probability under this distribution as
\begin{equation}
    p_{\rm tail}(p_V^\star)= \min\!\left(\int_0^{p_V^\star} p_{2R}(p_V)dp_V,\int_{p_V^\star}^\infty p_{2R}(p_V) dp_V\right),
\end{equation}
so that both unusually low and unusually high albedos are treated as anomalous. We then transform the rarity into a standardized score,
\begin{equation}
    s_{\rm albedo} = 1 - 2p_{\rm tail}(p_V^\star)
\end{equation}
which lies in $[0,1]$ and approaches unity for increasingly unlikely albedos. The albedo metric can be defined using this procedure and it allows $D$ to be interpreted as a rarity score based on the empirically established albedo distribution.
\\
\\
Applying this framework to the three known ISOs proceeds as follows. For 2I/Borisov, the observed albedo is consistent with the dark mode of the Solar System distribution ($p_V\sim0.04$), giving a small tail probability and hence $D\approx 0.10$ consistent with a normal cometary body. For 1I/`Oumuamua, albedo estimates cluster around $p_V\sim0.10$-0.15, which sits between the two Rayleigh components and is somewhat rarer, yielding $s_{\rm albedo}$ near $0.7$ and thus $D\approx 0.70$. For 3I/ATLAS, the derived effective radius and brightness imply an unusually high albedo compared to cometary populations, yielding $s_{\rm albedo}$ of order $0.6$ and thus $D\approx 0.60$. 
\\
\\
The trajectory/targeting anomaly $E$ is naturally constructed from the arrival geometry probability $p$ under an isotropic arrival model. A suitable mapping uses the log-probability
\begin{equation}    E=\operatorname{clamp}\!\left(\frac{-\log_{10}(p)}{X},0,1\right),
\end{equation}
where $X$ is a tunable scale constant (e.g. $X=3$ so that $p=10^{-3}$ maps close to $E=1$). We can also adopt the electromagnetic-signal score $F$ to range from $0$ (no detected artificial signals above background) to $1$ (narrowband, persistent, high signal-to-noise ratio signals inconsistent with known astrophysical emission processes). In practice $F$ can be modeled as a monotonic function of signal-to-noise and narrowness as follows,
$$
F=\operatorname{clamp}\left( \frac{{\rm SNR}_\text{narrow}/({\rm SNR}_\text{narrow}+K_F)}{1+\delta_{\rm astro}},0,1\right),
$$
where ${SNR}_\text{narrow}$ measures narrowband signal significance, $K_F$ a calibration constant and $\delta_{\rm astro}$ penalizes signals plausibly of natural astrophysical origin. For the present framework we may treat $F$ as effectively binary in many cases, with $F\approx 0$ or $F\approx 1$.
\\
\\
The operational/behaviour score $G$ encodes evidence for maneuvers, deployed sub-objects, or observation-responsive behavior. One may construct $G$ from an ensemble of indicators and normalize into $[0,1]$ using logistic or rational maps. Finally, the impact-risk factor $H$ is optional for the detection phase classification but required to map confirmed technology to Loeb levels 8-10, which can be a normalized function of impact probability and kinetic energy.
\\
\\
With normalized metrics in hand we construct a single composite Loeb score $S \in [0,1]$ that aggregates both the individual contributions of each observable feature and possible synergies between them. The simplest form is a weighted linear sum over the metrics,
\begin{equation}
    S_{lin}=\sum_{i\in\{A,B,C,D,E,F,G\}} w_i m_i, 
\end{equation}
where $m_i$ denotes the normalized metric values and the weights $w_i$ satisfy $w_i \ge 0$ and $\sum_i w_i = 1$. One also needs a form of an anomaly counting rule, and so we allow for non-linear reinforcement when distinct anomalies appear together. This is achieved by adding pairwise interaction terms,
\begin{equation}
    S = \sum_i w_i m_i + \sum_{i<j} w_{ij}m_i m_j,
\end{equation}
where $w_{ij}$ quantifies the strength of the interaction between metrics $m_i$ and $m_j$. In this way, a combination such as an unusual spectral signature appearing simultaneously with an extreme albedo value can raise the score more than either feature would on its own, reflecting the notion that multiple independent anomalies provide stronger evidence than a simple linear sum would capture. The interaction weights can be constrained so that only physically motivated pairs contribute, keeping the scheme transparent and interpretable.
\\
\\
As a starting recommendation for the linear weights, we adopt
$$
w_A=0.30,\quad w_B=0.15,\quad w_C=0.15,\quad w_D=0.10,\quad w_E=0.15,\quad w_F=0.10,\quad w_G=0.05,
$$
which reflect the judgment that non-gravitational acceleration is particularly diagnostic while still preserving substantial weight for spectral, geometric, and trajectory indicators. These weights sum to unity, and the additional pairwise terms allow the framework to capture the fact that coincident anomalies are often more significant than their parts considered in isolation.
\\
\\
Some observational data must also act as hard trigger overrides in accord with the Loeb Scale's logic. If a hard trigger observation is made, then the Loeb level should be immediately escalated regardless of the value of $S$. Representative override rules can be as follows:
\begin{enumerate}
\item If narrowband, persistent electromagnetic signals are confirmed and withstand community review, they set a minimum provisional Loeb level of 6 (suspected active technology) or higher depending on corroborating evidence.
\item If deployed sub-objects or clear controlled maneuvers are directly observed (like motion incompatible with gravity and outgassing and reproducible across epochs), they set a minimum provisional Loeb level of 6 and consider immediate elevation to level 8 after confirmation.
\item If multiple independent confirmation lines (morphology, composition inconsistent with cosmic ray exposure, persistent electromagnetic signals) establish artificial construction beyond reasonable doubt, they assign level 8 (confirmed technology with no impact) and apply the impact-differentiation protocol using $H$ to determine levels 9 or 10 if a collision trajectory with regional or global consequences is present.
\end{enumerate}
These override rules ensure that exceptionally decisive evidence is not diluted by averaging.
\\
\\
To map $S$ to discrete Loeb integer levels, we adopt a thresholding scheme that preserves the conceptual placement of the critical Level 4 threshold as the entry point for formal technosignature consideration \cite{eldadi2025loeb}. A suggested mapping is as follows:
$$
\begin{aligned}
S &<0.20 &\Longrightarrow&\ \text{Loeb Level }0,\\
0.20\le S &<0.35 &\Longrightarrow&\ \text{Loeb Level }1,\\
0.35\le S &<0.50 &\Longrightarrow&\ \text{Loeb Level }2,\\
0.50\le S &<0.60 &\Longrightarrow&\ \text{Loeb Level }3,\\
0.60\le S &<0.70 &\Longrightarrow&\ \text{Loeb Level }4,\\
0.70\le S &<0.80 &\Longrightarrow&\ \text{Loeb Level }5,\\
0.80\le S &<0.90 &\Longrightarrow&\ \text{Loeb Level }6,\\
0.90\le S &<0.95 &\Longrightarrow&\ \text{Loeb Level }7,\\
0.95\le S &<0.98 &\Longrightarrow&\ \text{Loeb Level }8,\\
0.98\le S &<0.995 &\Longrightarrow&\ \text{Loeb Level }9,\\
S &\ge 0.995 &\Longrightarrow&\ \text{Loeb Level }10
\end{aligned}
$$

The numerical boundaries are tunable with the important conceptual features being that Level 4 corresponds to $S$ above a high but not extreme value (here $\sim 0.60$) and that the highest levels are reserved for near-certain confirmation and impact risk. Measurement uncertainty on each metric propagates to uncertainty in $S$ and so let each metric $m_i$ have an estimated standard error $\sigma_{m_i}$. Under the linear combination approximation (neglecting covariances for a first-order estimate), the standard error of $S$ is
\begin{equation}
    \sigma_S=\sqrt{\sum_i (w_i \sigma_{m_i})^2},
\end{equation}
In general, it may be best to propagate measurement distributions via Monte Carlo sampling if metric errors are non-Gaussian or covariances are important. One should report the composite as $S\pm\sigma_S$ and when communicating to stakeholders, present the probability mass function over discrete Loeb levels obtained by sampling the uncertainty distribution and applying the $S\mapsto\text{Level}$ mapping to each draw. The framework must be calibrated where calibration steps can include choosing $a_{\rm ref},K_B,K_D,R_{\max},X$ and the weights $w_i$ by fitting to historical cases and simulated injections.
\\
\\
\section{Examples}
We now illustrate the framework with worked numerical examples based on the qualitative and quantitative summaries available for the three currently discussed ISOs: 1I/`Oumuamua, 2I/Borisov, and 3I/ATLAS. The numerical values used here are illustrative and chosen to reproduce the qualitative assessments in \cite{eldadi2025loeb}.
We can adopt the following constants and weights for these examples: $a_{\rm ref}=10^{-6}\ \mathrm{m,s^{-2}}$, $R_{\max}=10$, $X=3$, $K_B=1$, $K_D=1$, 
We also adopt these linear weights throughout: 
\begin{equation}
    w_A=0.30,\quad w_B=0.15,\quad w_C=0.15,\quad w_D=0.10,\quad w_E=0.15,\quad w_F=0.10,\quad w_G=0.05.
\end{equation}
We introduce a sparse set of symmetric pairwise interaction weights $w_{ij}=w_{ji}$ (all other $w_{ij}=0$). For clarity, we list only the nonzero $w_{ij}$ used in these illustrative calculations
$$
\begin{aligned}
&w_{AB}=0.050,\quad w_{AC}=0.030,\quad w_{BC}=0.030,\\
&w_{AE}=0.040,\quad w_{BE}=0.030,\quad w_{AD}=0.010,\\
&w_{BD}=0.010,\quad w_{DE}=0.020.
\end{aligned}
$$
These values are small by construction (typically a few percent) so the linear terms remain the dominant contribution but pairwise synergies produce the modest boosts desired when multiple independent anomalies co-occur. In practice, one should regularize these $w_{ij}$ and restrict the allowed pairs to physically motivated combinations. We again note here that the choices above are illustrative and were chosen so the final discrete Loeb assignments reproduce the classifications for the three examples \cite{eldadi2025loeb}.
\\
\\ 
{\bf Example 1: 1I/`Oumuamua}. The reported non-gravitational acceleration is $a_{\rm obs}=4.92\times 10^{-6}\ \mathrm{ms^{-2}}$ and for this case we compute
\begin{equation}
    A_{\rm raw}=\log_{10}\!\left(\frac{4.92\times10^{-6}}{10^{-6}}\right)=\log_{10}(4.92)\approx 0.692,
\end{equation}
and thus we have
\begin{equation}
    A=\operatorname{clamp}\!\left(\frac{0.692+2}{4},0,1\right)=\frac{2.692}{4}\approx 0.673.
\end{equation}
We assign spectral anomaly $B\approx 0.80$ to reflect the lack of detected volatiles, shape anomaly $C\approx 1.00$ to reflect extreme inferred aspect ratio, albedo/weathering anomaly $D\approx 0.70$ for surface/compositional oddities, trajectory score $E\approx 0.10$ for no strong targeting, and $F=0$, $G=0$. 
The linear sum is
$$
\begin{aligned}
S_{\rm lin} &= w_A A + w_B B + w_C C + w_D D + w_E E + w_F F + w_G G \\
&= 0.30\times 0.673 + 0.15\times 0.80 + 0.15\times 1.00 + 0.10\times 0.70 + 0.15\times 0.10 + 0.10\times 0 + 0.05\times 0 \\
&\approx 0.2019 + 0.1200 + 0.1500 + 0.0700 + 0.0150 + 0 + 0 \\
&= 0.5569.
\end{aligned}
$$
The nonzero interaction products (with our chosen $w_{ij}$) and their contributions are
$$
\begin{aligned}
m_A m_B &= 0.673\times 0.80 = 0.5384 &\Rightarrow&\quad w_{AB} m_A m_B = 0.050\times 0.5384 \approx 0.0269,\\
m_A m_C &= 0.673\times 1.00 = 0.6730 &\Rightarrow&\quad w_{AC}\,m_A m_C = 0.030\times 0.6730 \approx 0.0202,\\
m_B m_C &= 0.80\times 1.00 = 0.8000 &\Rightarrow&\quad w_{BC}\,m_B m_C = 0.030\times 0.8000 = 0.0240,\\
m_A m_E &= 0.673\times 0.10 = 0.0673 &\Rightarrow&\quad w_{AE}\,m_A m_E = 0.040\times 0.0673 \approx 0.0027,\\
m_B m_E &= 0.80\times 0.10 = 0.0800 &\Rightarrow&\quad w_{BE}\,m_B m_E = 0.030\times 0.0800 = 0.0024,\\
m_A m_D &= 0.673\times 0.70 = 0.4711 &\Rightarrow&\quad w_{AD}\,m_A m_D = 0.010\times 0.4711 \approx 0.0047,\\
m_B m_D &= 0.80\times 0.70 = 0.5600 &\Rightarrow&\quad w_{BD}\,m_B m_D = 0.010\times 0.5600 = 0.0056,\\
m_D m_E &= 0.70\times 0.10 = 0.0700 &\Rightarrow&\quad w_{DE}\,m_D m_E = 0.020\times 0.0700 = 0.0014.
\end{aligned}
$$
Summing the interaction contributions yields
$$
S_{\rm int}\approx 0.0269+0.0202+0.0240+0.0027+0.0024+0.0047+0.0056+0.0014\approx 0.0879.
$$
The total composite score is
$$
S_{\rm `Oumuamua}=S_{\rm lin}+S_{\rm int}\approx 0.5569+0.0879=0.6448.
$$
Using the mapping to Loeb levels, $0.60\le S<0.70$ implies Loeb Level 4, consistently with \cite{eldadi2025loeb}.
\\
\\
{\bf Example 2: 2I/Borisov}. We assign conservative values consistent with a classical comet: $A\approx 0.25$, $B\approx 0.00$, $C\approx 0.176$ (mild lightcurve/aspect ratio), $D\approx 0.10$, $E\approx 0$, $F=0$, $G=0$. The linear sum is
$$
\begin{aligned}
S_{\rm lin} &= 0.30\times 0.25 + 0.15\times 0.00 + 0.15\times 0.176 + 0.10\times 0.10 + 0.15\times 0 + 0 + 0 \\
&\approx 0.0750 + 0 + 0.0264 + 0.0100 = 0.1114.
\end{aligned}
$$
Interaction products that could be nonzero in principle are small because many $m_i$ vanish or are very small. The only modest contributions among our chosen pairs are
$$
m_A m_C = 0.25\times 0.176 = 0.0440 \quad\Rightarrow\quad w_{AC}\,m_A m_C = 0.030\times 0.0440 \approx 0.0013,
$$
and
$$
m_B m_D = 0.00\times 0.10 = 0 \quad\Rightarrow\quad w_{BD}\,m_B m_D = 0.
$$
Other pairwise products involving $B,F,G,E$ are zero and so
$$
S_{\rm int}\approx 0.0013,
$$
and the total score is
$$
S_{\rm Borisov}\approx 0.1114+0.0013=0.1127.
$$
This remains well below the $S<0.20$ threshold and thus maps to Loeb Level 0, again consistently with the classification \cite{eldadi2025loeb}.
\\
\\
{\bf Example 3: 3I/ATLAS}. We assign $A\approx 0.70$ (moderate non-gravitational/compositional anomaly), $B\approx 0.80$ (spectral paradox: dust but no gas), $C\approx 0.30$ (moderate lightcurve amplitude), $D\approx 0.60$ (size/albedo anomaly) and use the trajectory probability $p\approx 0.002$ to compute
$$
-\log_{10}(p)\approx 2.699,\quad E=\operatorname{clamp}\!\left(\frac{2.699}{3},0,1\right)\approx 0.900.
$$
We set $F=0$, $G=0$ and so The linear sum is
$$
\begin{aligned}
S_{\rm lin} &= 0.30\times 0.70 + 0.15\times 0.80 + 0.15\times 0.30 + 0.10\times 0.60 + 0.15\times 0.90 + 0 + 0 \\
&= 0.21 + 0.12 + 0.045 + 0.06 + 0.135 = 0.57.
\end{aligned}
$$
The nonzero pairwise products and their contributions (with our chosen $w_{ij}$) are
$$
\begin{aligned}
m_A m_B &= 0.70\times 0.80 = 0.5600 &\Rightarrow&\quad w_{AB}\,m_A m_B = 0.050\times 0.5600 = 0.0280,\\
m_A m_E &= 0.70\times 0.90 = 0.6300 &\Rightarrow&\quad w_{AE}\,m_A m_E = 0.040\times 0.6300 = 0.0252,\\
m_B m_E &= 0.80\times 0.90 = 0.7200 &\Rightarrow&\quad w_{BE}\,m_B m_E = 0.030\times 0.7200 = 0.0216,\\
m_A m_D &= 0.70\times 0.60 = 0.4200 &\Rightarrow&\quad w_{AD}\,m_A m_D = 0.010\times 0.4200 = 0.0042,\\
m_B m_D &= 0.80\times 0.60 = 0.4800 &\Rightarrow&\quad w_{BD}\,m_B m_D = 0.010\times 0.4800 = 0.0048,\\
m_D m_E &= 0.60\times 0.90 = 0.5400 &\Rightarrow&\quad w_{DE}\,m_D m_E = 0.020\times 0.5400 = 0.0108.
\end{aligned}
$$
Summing these interaction contributions yields
$$
S_{\rm int}\approx 0.0280 + 0.0252 + 0.0216 + 0.0042 + 0.0048 + 0.0108 \approx 0.0946.
$$
The total composite score is therefore
$$
S_{\rm ATLAS}=S_{\rm lin}+S_{\rm int}\approx 0.57 + 0.0946 = 0.6646.
$$
Using the thresholds above, $0.60\le S<0.70$ corresponds to Loeb Level 4, again consistently with the conclusions in \cite{eldadi2025loeb} 
\\
\\
Uncertainty estimates for the examples will follow a propagation formula, considering if each metric $m_i$ has estimated standard error $\sigma_{m_i}$ and if we treat the interaction weights and metric errors as independent in a first approximation, the variance of $S$ under a linearized error model is
\begin{equation}
    \sigma_S^2 \approx \sum_i \left(w_i + \sum_{j\ne i} w_{ij} m_j\right)^2 \sigma_{m_i}^2,
\end{equation}
so that
\begin{equation}
    \sigma_S \approx \sqrt{\sum_i \left(w_i + \sum_{j\ne i} w_{ij} m_j\right)^2 \sigma_{m_i}^2 }.
\end{equation}
In practice, Monte Carlo sampling of the posterior distributions for the $m_i$ with the full non linear expression for $S$ is preferable and the above formula serves as a first order approximation. Presenting $S\pm\sigma_S$ and the Monte Carlo derived probability mass across Loeb levels is recommended for transparent communication when concerning decisions of global impact on Earth.
\\
\\
\section{Conclusions}
The proposed framework for quantifying the Loeb Scale is modular in the sense that the metric definitions, transforms, calibration constants, and weights are all explicitly configurable and can be refined by a working group of the International Astronomical Union or an equivalent community process. Hard-trigger override rules ensure that decisive evidence is acted upon promptly and the anomaly count bonus encourages detection teams to value corroboration across independent data types (trajectory, spectrum, geometry, electromagnetic signals) without letting counting alone dominate the weighted evidence.
\\
\\
Finally, governance and policy decisions follow naturally from this mapping. For $S$ in the range corresponding to Loeb Level 4, the framework suggests enhanced global observational campaigns, prioritized telescope time, and transparent rapid data release. For Levels 5-7, the framework recommends escalating coordination with agencies responsible for planetary defense, and for Levels 8-10 it prescribes emergency protocols analogous to those used in planetary-impact scenarios but adapted to the additional complications related to artificial causation. The mathematics above provides a reproducible, transparent, and tunable bridge between raw measurements and the Loeb Scale integer classification, allowing the astronomical community to communicate consistently and to exercise preparedness while avoiding premature actions.
\\
\\
It is important to emphasize that the Loeb Scale is conceptually independent of the Drake Equation as the Drake Equation was originally formulated as a probabilistic tool to estimate the number of contemporaneous radio communicating civilizations in the Milky Way \cite{dr1vakoch2015drake,dr2maccone2012statistical,dr5drake2013reflections}. The terms of the Drake equation explicitly encode astrophysical, biological, and sociological factors that govern the occurrence rate of detectable electromagnetic signals, and are therefore tied to the question of how many civilizations might be actively transmitting detectable signals \cite{dr3burchell2006w,dr4glade2012stochastic,dr6prantzos2013joint}. In contrast, the Loeb Scale is not a measure of occurrence rates but a classification framework for interpreting the technosignature significance of physical artifacts that are detected near-Earth, regardless of when or by whom they were produced. 
\\
\\
This distinction has practical implications as one may imagine scenarios in which the Galaxy contains no radio emitting civilizations at present, either because advanced societies perished by now or because they no longer engage in radio communication. Under these circumstances, it could still be possible that long-lived artifacts such as probes, fragments, or derelict spacecraft are present for billions of years, thus entering the Solar System as interstellar objects. The Loeb Scale is designed precisely for such cases as it quantifies the anomaly level and possible artificial nature of nearby physical objects without making any assumptions about the present-day activity of their creators. Thus, the Loeb Scale complements, but does not depend on, the Drake framework, because it deals with surviving physical technosignatures rather than transient electromagnetic signals.
\\
\\
\section*{Acknowledgements}
OT was supported in part by the Vanderbilt Discovery Alliance Fellowship. AL was supported in part by the Galileo Project and the Black
Hole Initiative.

\bibliography{references}
\bibliographystyle{unsrt}

\end{document}